# GaAs as a bright cryogenic scintillator for the direct dark matter detection

S. Vasiukov, F. Chiossi, C. Braggio, G. Carugno, F. Moretti, E. Bourret and S. Derenzo

*Abstract*— The optical and scintillation properties of undoped, Si-doped, and Si, B co-doped GaAs samples were studied. The light yield specification and X-ray luminescence of GaAs over a wide IR region by using Si and InGaAs photodetectors are presented. The undoped GaAs demonstrated a narrow emission band at 838 nm (1.48 eV) and a low light output of about 2 ph/keV. The GaAs:Si is characterized by three broad luminescence bands at 830 nm (1.49 eV), 1070 nm (1.16 eV), 1335 nm (0.93 eV) and a light output of about 71 ph/keV. In the case of GaAs:(Si, B), four luminescence bands at 860 nm (1.44 eV), 930 nm (1.33 eV), 1070 nm (1.16 eV) and 1335 nm (0.93 eV) were observed. A high light yield higher than 125 ph/keV was estimated for the co-doped sample. The applications of low energy gap semiconductors as cryogenic scintillators for particle detection are discussed. The GaAs is a promising crystal for a cryogenic scintillator for the dark matter (DM) detection.

## I. Introduction

Semiconductor-based scintillation materials are of great importance for science and technology and have a great history, starting in 1909 with the well-known Rutherford gold foil experiments, using ZnS to detect alpha particles. Nowadays the semiconductor applications as scintillators are no less relevant. According to the last theoretical estimation and experimental results, low band gap materials can be effective for direct detection of dark matter (DM) in the MeV range [1-4]. Such light masses may be detected by dark matter-electron scattering. Besides, the scintillator light output increases with decreasing of band gap [5-9]. In the case of GaAs crystal, the maximum value of this parameter should be about 280 ph/keV.

Gallium Arsenide (GaAs) was proposed as a promising cryogenic detector and its scintillation characteristics were determined in the previous work [10]. These measurements show that silicon and boron doped GaAs is a promising cryogenic scintillator for DM particle detection in the MeV/c$^2$ mass range in that it can be grown as large, high-quality crystals, has good scintillation luminosity (43 ph/keV), and potentially no afterglow. Two emission bands at 850 nm and 930 nm were observed and a silicon photodetector was used to determine the light output of this emission. Generally, Silicon photodetectors are sensitive to long-wavelength photons up to 1000 nm. However, in GaAs:Si luminescence at 1050 nm and 1300 nm was observed in previous works [11-12]. Thus, for a more accurate determination of the light output of GaAs:(Si, B) it is necessary to expand the wavelength range and use an InGaAs photodetector with sensitivity in the range from 800 nm to 1700 nm.

This paper presents the results of light yield (LY) estimation of undoped GaAs, GaAs:(Si) and co-doped GaAs:(Si, B) at low temperature by using Si and InGaAs photodetectors.

## II. Experimental Details

### A. GaAs Samples

Table I lists the GaAs samples used in the measurements. All samples were grown by vertical gradient freeze (VGF) method. They are in the form of wafers with 450-500 μm thicknesses. Gallium arsenide samples were one side polished wafers.

TABLE I
GALLIUM ARSENIDE SAMPLES USED IN THIS WORK

| Sample | Supplier | Si ppm | B ppm | free carriers/cm$^3$ | Notes |
|---|---|---|---|---|---|
| GaAs:(Si, B) | AXT Inc | 8.9 | 9.7 | $5.5 \times 10^{17}$ | No 13316 see [10] |
| GaAs:Si | Wafer technology LTD | 6 | - | $2.5 \times 10^{18}$ | |
| GaAs (pure) | Wafer technology LTD | - | - | $3 \times 10^7$ | |

### B. Cryostat

All measurements were performed at 10 K and a closed-type Leybold-Heraeus helium refrigerator cryostat was used.

### C. X-ray luminescence

The radioluminescence spectra were measured using spectrometers based on Si and InGaAs arrays. The Thorlabs CCS 175 spectrometer uses a Silicon CCD and provides detection in the 500-1000 nm range. The Ocean Optics NIRQuest 512 spectrometer uses an InGaAs 512 element linear

S. Vasiukov (vasyukov_sergey@yahoo.com), F. Chiossi, C. Braggio, G. Carugno are with the Department of Physics and Astronomy, University of Padua and Istituto Nazionale Fisica Nucleare, sez. Padova via F-Marzolo 8, I-35131 Padua, Italy

F. Moretti, E. Bourret are with the Materials Sciences Division, Lawrence Berkeley National Laboratory, 1 Cyclotron Road, Berkeley, California 94720, USA

S. Derenzo is with the Molecular Biophysics and Integrated Bioimaging Division, Lawrence Berkeley National Laboratory, 1 Cyclotron Road, Berkeley, California 94720, USA array to cover the wavelength range from 900-1700 nm. Correction curves for both spectrometers were provided by the manufacturers and were used in data processing.

*D. Light Yield Estimation*

To estimate the light yields, the samples were excited by X-rays. The Tungsten anode OCX/70-G tube (Voltage up to 85 kV) by Compagnia Elettronica Italiana (C.E.I.) was used as an X-ray source. The high voltage power supplies HCP 70–65000 by FuG Elektronik GmbH ($V_{max}$=65 kV, $I_{max}$=100 mA) provided power to the X-ray tube.

The X-ray induced emission is analyzed in a wide wavelength band covering the range from visible up to near infrared. PIN photodetectors, based on Si and InGaAs were used to determine the light output of the samples. The first one is Hamamatsu Si PIN photodiode S11499-01 which offers enhanced near-infrared sensitivity in 360-1140 nm range. And the second PIN photodiode is InGaAs Hamamatsu G12180-250A with two-stage thermoelectrically-cooling which provide low noise background in 900-1700 nm spectral response. Both photodiodes had a 5 mm diameter photosensitive area. The photodetector current was measured by KEITHLEY auto ranging picoammeter 485 (current range 2 nA-2 mA, resolution up to 0.1 pA).

Simplified scheme of the experimental apparatus shown in Fig. 1. The X-ray beam passes through the thin Teflon plate (1), is collimated by aluminum disk (2) and falls on crystalline sample (3). The sample (3) is mounted by the polished side to X-ray source and between disk (2) and copper masks (4) with 4.3 mm hole on the copper holder (5) for better thermal contact. The X-ray losses before they hit the sample are insignificant since Teflon and aluminum are 4 mm and 0.3 mm thick, respectively. At the same time, the sample thickness is sufficient to absorb 99% of the total incident radiation. The X-ray induced light coming out of the sample and exits out of the cryostat (6) through the sapphire window (7) with 5 mm length and 25 mm in diameter.

The light then passes through a quartz light guide (8) wrapped with several layers of a reflective Teflon film and is detected by a photodetector (9), which, depends on the task, is an appropriate PIN photodiode or spectrometer (see section II C).

The PIN photodetectors are sensitive to direct X-ray radiation that is why the quartz light guide (100×10 mm, shown with a break in Fig. 1) and lead sheets (not shown) were used to eliminate the spurious signal by distancing the photodetector from the noise source when the useful signal from the sample is stored. By this way, the direct effect of the X-ray source on the photodetector was completely eliminated. The dark current of the PIN photodiode was about 0.01 nA (InGaAs photodiode) and 0.1 nA (Si photodiode) when the X-ray tube was switched on. It is important to note that in case of InGaAs photodiode the thermoelectric cooling made it possible to obtain a smaller dark current in comparison with Si photodiode which was used at room temperature (without cooling).

The photodiode signals were calibrated using the measurement of 43 ph/keV over the 800-970 nm band for the same GaAs:(Si, B) sample reported in Reference [10]. The photodiode signals observed using a FES950 filter that blocked wavelengths above 950 nm, and a correction for the estimated 25.6% of the intensity in the 950-970 nm region. This calibration was used in the following sections to estimate the light output in the different emission bands for the different samples in Table I. Thorlabs shortpass (FES) and longpass (FEL) filters were used to selectively measure the photoresponse at different wavelength ranges. This made it possible to isolate regions of a spectrum and estimate the fraction of emission bands separately in a wide range. The relative light output was calculated taking into account the filter transmission and the spectral sensitivity of photodetectors.

### III. RESULTS AND DISCUSSION

*A. GaAs Emission Under X-ray Excitation*

The measurement of X-ray luminescence spectra is an important element of this investigation. The obtained data allow us to determine the distribution of all the radiative centers in the range of 800 – 1700 nm and estimate the contribution of each component. The results of these experiments are presented in Fig. 2.

Only one narrow emission band at 838 nm (A band) is observed in the undoped GaAs crystals at low temperature. Three emission bands at 830 nm (A band), 1070 nm (band C) and weak band at ~1300 nm (band D) are observed in GaAs:Si. The A band is considerably wider and asymmetrical in this case. And four luminescence bands at 830 nm (band A), 930 nm (band B), 1070 nm (band C), 1335 nm (band D) are clearly manifested in co-doped GaAs:(Si, B) crystal. It should be noted that the intensity of the A band is very low compared to other samples. The obtained data are consistent with those reported in the literature [10-18]. The emission center parameters and their supposed nature are shown in Table II. The short wavelength emission is attributed to the band-to-band transition (~817 nm), excitons relaxation or shallow levels (defects) in case of undoped GaAs. The band-to-acceptor (~835 nm) and donor-to-acceptor (~870 nm) transition and other occur in doped GaAs crystals. But in the framework of this work, such features will not be considered. For simplicity, these emission peaks were collected in one group – band A. The nature and specific of these short wavelength emissions can explains the position and asymmetric shape of band A in GaAs:Si and GaAs:(Si, B) (see Fig. 2). As can be seen from the experimental and literature data, the shape, intensity and peak position of the bands in 800-900 nm region in GaAs crystals drastically depends on purity, type of dopants and defects [11-15]. And it also depends, as shown in Reference [13], on the type of conductivity.

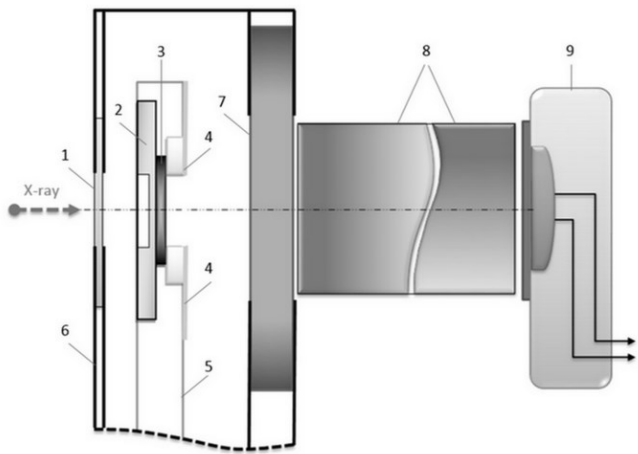

Fig. 1. The scheme of experimental equipment for light yield estimation. The description is in the text.

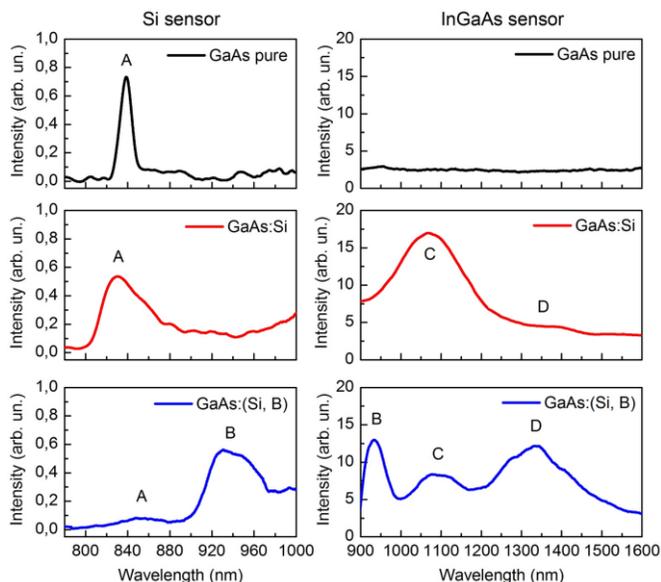

Fig. 2. The X-ray luminescence spectra of undoped GaAs, GaAs:Si and GaAs:(Si, B) crystals obtained by Si (left column) and InGaAs (right column) CCDs at 10K.

The transitions from shallow silicon donors to boron acceptors on an arsenic site leads to emissions at 930 nm (band B) [10, 16, 17].

The energy relaxation on gallium vacancy-donor center leads to emission at ~1070 nm [18]. This center (band C) is a gallium vacancy ($V_{Ga}$) bound to a donor (like Si, Ge, Sn, C, S, Se, Te). The peak position of band C depends on the type of activator impurity.

The nature of the broad band D at ~1335 nm is associated with defects or complex centers like ($Si_{Ga} + V_{Ga} + Si_{Ga}$) or ($Si_{Ga} + Si_{As}$) and other [11, 12]. It is important to note that the formation of long-wavelength centers at ~1377 nm was observed in undoped GaAs crystals with neutron dose. The intensity of band D increases with neutrons dose rate increasing, which indicates the important role of vacancies in the formation of this emission center [12].

### B. Light Yield Estimation

The photocurrent of all samples under X-ray was measured using the InGaAs or Si photodiodes with or without filters and shown on Fig. 3. Table III compiles the total light output and the contribution of each emission band separately which were obtained in this work based on PIN detectors photocurrent.

The obtained results indicate a very high efficiency of the scintillation process in activated crystals. It is seen that the LY of the GaAs:(Si, B) is very high: 125 ph/keV. The Si doped GaAs crystal has LY=71 ph/keV. It is important to note that about 65% of the total output of activated GaAs is contained in the region of 1000-1700 nm, i.e. bands C and D. This feature is dictated by the specificity of the radiative relaxation of energy in the co-doped GaAs crystal. The results show how important the addition of the acceptor boron is. Also, this indicates the need to use photo detectors that are sensitive in a fairly wide range (like Ge, InGaAs). Silicon-based photodiodes are only partially suitable as photodetectors, since 35 to 60% of the light is lost in case of GaAs:(Si, B) and GaAs:Si respectively.

TABLE II
THE PARAMETERS OF RADIATIVE CENTERS IN GaAs CRYSTALS

| Band | λ, nm | E, eV | Comments |
|---|---|---|---|
| A | 800-870 | 1.42-1.54 | Transition from CB to VB, exciton and shallow levels (defects) |
| B | 933 | 1.33 | The boron in arsenide site (BAs center) |
| C | 1050-1090 | 1.14-1.18 | Center like ($Si_{Ga} + V_{Ga}$) |
| D | 1305-1335 | 0.93-0.95 | Complex center like ($Si_{Ga} + V_{Ga} + Si_{Ga}$) or ($Si_{Ga} + Si_{As}$) |

It should be noted that the experimental approach used to determine the light output does not take into account the refractive index. GaAs has a refractive index of 3.55 at 930 nm and 3.4 at 1300 nm. Since the scintillation light is emitted isotropically, only a small fraction is able to exit the crystal directly. In this experiment, scattering within the crystal or on the roughened surface allowed a large fraction of the light to exit. A coating with a graded index of refraction will allow

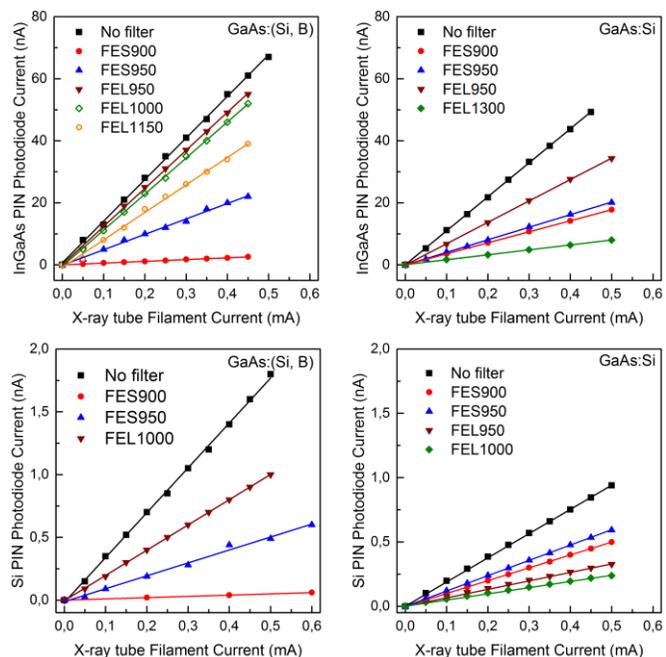

Fig. 3. The InGaAs (Top) and Si (Bottom) PIN photodiodes current dependence on X-ray filament current with/without the different shortpass (FES) and longpass (FEL) optical filters obtained on the GaAs:(Si, B) (Left) and GaAs:Si (Right) sample at 10 K.

TABLE III
THE LIGHT OUTPUT OF GALLIUM ARSENIDE SAMPLES FOR DIFFERENT SPECTRAL REGIONS AND PHOTODETECTORS

| Crystal | PD | Region (nm) | Bands | LY (%) | LY (ph/keV) |
|---|---|---|---|---|---|
| GaAs:(Si, B) | InGaAs | 800-900 | A | 1 | 1.2 |
|  |  | 900-1000 | B | 33 | 42 |
|  |  | 1000-1150 | C | 28 | 35 |
|  |  | 1150-1700 | D | 38 | 47 |
|  |  | 800-1700 | Total | 100 | 125.2 ± 6 |
| GaAs:(Si, B) | Si | 800-900 | A | 2 | 1.6 |
|  |  | 900-1000 | B | 48 | 39 |
|  |  | 1000-1140 | C | 49 | 40 |
|  |  | 800-1140 | Total | 100 | 80.6 ± 4 |
| GaAs:Si | InGaAs | 800-950 | A | 35 | 25 |
|  |  | 950-1300 | C | 51 | 36 |
|  |  | 1300-1700 | D | 14 | 10 |
|  |  | 800-1700 | Total | 100 | 71 ± 4 |
| GaAs:Si | Si | 800-950 | A | 60 | 25 |
|  |  | 950-1140 | C | 40 | 17 |
|  |  | 800-1140 | Total | 100 | 42 ± 2 |
| GaAs pure | Si | 800-1140 | Total | 100 | 2 ± 0.5 |

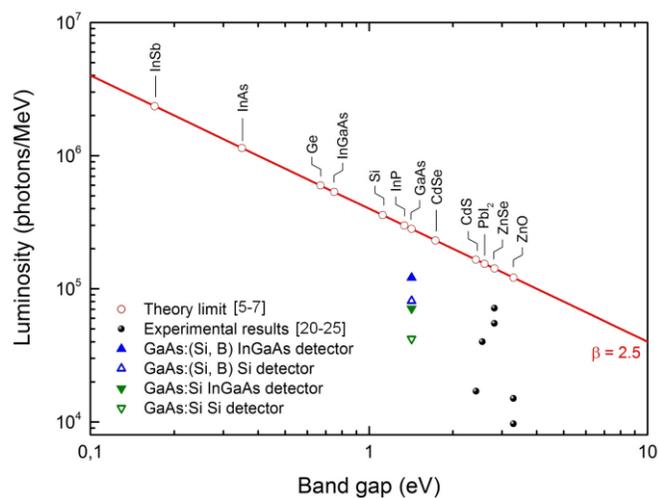

Fig. 4. The dependence of the LY on the value of the material's band gap.

essentially all of the scintillation photons to escape directly [19].

### C. Semiconductors as promising bright scintillators

The well-known rule of the scintillation limit of light yield $LY < 1\ MeV/(2.5 \times E_g)$ with $E_g$ the scintillator band gap, which was proposed in Reference [10-12], makes it easy to estimate the maximum brightness of the scintillation material. It is important to note that the light output should become very large for vanishing bandgaps (see Fig. 4).

Despite the high light output and, as a consequence of the potentially high efficiency, semiconductor scintillators have several important features. Generally, the main obstacles to widespread use of the low bandgap cryogenic scintillator are the need of low-temperatures and the lack of good NIR photodetectors.

Firstly, by cooling down to cryogenic temperature the thermal quenching of the light emission in a semiconductor decreases. However, it leads to complication of the experiment apparatus, because the cryostat chamber is necessary.

Secondly, for example, superconducting nanowire single photon detectors have high quantum efficiency over the 800-1700 nm range and very low dark pulse rates, but the largest reported devices are less than 1 mm$^2$. The similar situation with a low sensors area holds true for avalanche photodiodes based on Si, Ge, InGaAs.

In addition to potentially high light output and specific application conditions, semiconductor materials can be very versatile detectors. Doped GaAs and other semiconductors could be used as hybrid detectors. This means that cryogenic detectors might have a double (or triple) readout: it could be indeed possible to measure the heat delivered by the impinging particles in parallel with an electric and/or scintillation signal. This possibility would allow discrimination between events that release energy with different efficiency on different channels. As an example of such application of semiconductor materials can be given experiments in the framework of the COSINUS and LUCIFER projects. [26, 27].

Nevertheless, the data reported in this work demonstrate the great potential of scintillation applications of materials like GaAs, InP, InGaAs, etc. The scintillation characteristic study of narrow band gap materials will contribute to the development of the different science fields and will allow us to more accurately describe the processes of high-energy relaxations in solids.

### IV. CONCLUSION

High light yields of 125 ph/keV and 71 ph/keV are reported at low temperatures in GaAs:(Si, B) and GaAs:Si, respectively. With this, we re-opened low band gap materials as bright scintillators. In addition, using the example of a GaAs:(Si, B) crystal, the importance of donor and acceptor activation as a way to improve the scintillation efficiency is demonstrated. The performed studies demonstrate that the attractive scintillation properties of doped and co-doped GaAs crystals at 10 K are comparable to those of SrI$_2$:Eu and LaBr$_3$:Ce at room temperature.

These measurements show that n-type GaAs is a promising cryogenic scintillator for DM particle detection. GaAs crystals are commercially available and have very bright scintillation luminosity. However, more work is needed to optimize silicon and boron dopant concentrations, as well as IR antireflection coatings, in order to further improve the overall performance of GaAs crystal detectors and their coupling with suitable photodetectors.

To improve the overall energy sensitivity and resolution of the GaAs crystal we could exploit the cryogenic bolometric approach where both light and temperature rise are detected. Such direction is well established in DM search in the experiment such as CRESST and COSINUS.

ACKNOWLEDGMENT

The research is sponsored by Istituto Nazionale di Fisica Nucleare (INFN) within the AXIOMA project and the Office of Basic Energy Sciences of the U.S. Department of Energy at the Lawrence Berkeley National Laboratory under UC-DOE Contract No. DE-AC02-05CH11231. The authors wish to thank Fulvio Calaon, Mario Tessaro, Mario Zago, Massimo Rebeschini, Andrea Benato and Enrico Berto for the help with cryogenics and for the mechanical and electronic work on the experimental setup.